\documentclass[aps,pra,showpacs,twocolumn,superscriptaddress]{revtex4-1}
\usepackage{amsfonts}
\usepackage{amsmath}
\usepackage{amssymb}
\usepackage[colorlinks, citecolor=blue,linkcolor=red]{hyperref}
\usepackage{color}
\usepackage{float}
\usepackage{hyperref}
\usepackage{textcomp}
\usepackage[rightcaption]{sidecap}
\usepackage{subfigure}
\usepackage{graphicx}

\begin{document}

\title{Phase separation and hidden vortices induced by spin-orbit coupling in spin-1 Bose-Einstein condensates}
\author{Ji Li}
\affiliation{Beijing National Laboratory for Condensed Matter Physics,
Institute of Physics, Chinese Academy of Sciences,
Beijing 100190, China}
\affiliation{School of Physical Sciences,
University of Chinese Academy of Sciences, Beijing 100190, China}
\author{Yan-Mei Yu}
\affiliation{Beijing National Laboratory for Condensed Matter Physics,
Institute of Physics, Chinese Academy of Sciences,
Beijing 100190, China}
\affiliation{School of Physical Sciences,
University of Chinese Academy of Sciences, Beijing 100190, China}
\author{Kai-Jun Jiang}
\affiliation{State Key Laboratory of Magnetic Resonance and Atomic and Molecular Physics, Wuhan Institute of Physics and
Mathematics, Chinese Academy of Sciences, Wuhan 430071}
\affiliation{Center for Cold Atom Physics, Chinese Academy of Sciences, Wuhan 430071}
\author{Wu-Ming Liu}
\email{wliu@iphy.ac.cn}\affiliation{Beijing National Laboratory for Condensed Matter Physics,
Institute of Physics, Chinese Academy of Sciences,
Beijing 100190, China}
\affiliation{School of Physical Sciences,
University of Chinese Academy of Sciences, Beijing 100190, China}

\begin{abstract}
  We investigate phase separation and hidden vortices in spin-orbit coupled ferromagnetic
  Bose-Einstein condensates with rotation and Rabi coupling. The hidden vortices are invisible in density distribution but
  are visible in phase distribution, which can carry angular momentum like the
  ordinary quantized vortices. In the absence of the rotation,
  we observe the phase separation induced by the spin-orbit coupling and determine the entire
  phase diagram of the existence of phase separation. For the rotation case, in addition to the phase separation,
  we demonstrate particularly that the spin-orbit coupling can result in the hidden vortices and hidden vortex-antivortex pairs.
  The corresponding entire phase diagrams are determined, depending on the interplay of
  the spin-orbit coupling strength, the rotation frequency, and Rabi frequency, which reveals the critical condition of
  the occurrence of the hidden vortices and vortex-antivortex pairs. The hidden vortices here
  are proved to be long-lived in the time scale of experiment by the dynamic analysis. These findings not only provide
  a clear illustration of the phase separation in spin-orbit coupled spinor Bose-Einstein condensates,
  but also open a new direction for investigating the hidden vortices in high-spin quantum system.

\end{abstract}
 \maketitle

\section{INTRODUCTION}
Bose-Einstein condensates (BECs) in an optical dipole trap, known as spinor BECs, have been experimentally
realized and studied in a gas of $^{23}$Na and $^{87}$Rb atoms \cite{Stamper-Kurn1998,Barrett2001}, which brings
a new way to create topologically nontrivial structures \cite{Sadler2006,Ji2008,Khawaja2001,Choi2012,Kawaguchi2008,Hall2016,Ray2014}, owing to the
spin degrees of freedom and many possible order-parameter manifolds. Recently, miscibility-immiscibility
phase transition has drawn great interests in the spinor BECs \cite{Ho1996,Pu1998,Sabbatini2011,Zhou2008,Xi2011,Bandyopadhyay2017,Myatt1997,Matuszewski2009,Matuszewski2010,ocki2012}. Phase miscibility and separation have
been studied theoretically \cite{Ho1996,Pu1998,Sabbatini2011,Zhou2008,Xi2011,Bandyopadhyay2017} and experimentally \cite{Myatt1997} in the two-component BECs previously. The ground states of spin-1 BECs under a homogeneous magnetic field have also been investigated \cite{Matuszewski2009,Matuszewski2010,ocki2012}. It was shown that the
homogeneous magnetic field can induce the phase separation when spin-dependent interaction is antiferromagnetic \cite{Matuszewski2009,Matuszewski2010}. In particular, the recent spin-orbit (SO) coupling in quantum gases can be controlled and realized by using optical or magnetic fields,
which has played an important role in the recently discovered novel quantum states \cite{Lin2011,Wu2016,Huang2016,Wang2010,Su2012,Liu2012,Xu2011,Sinha2011,Hu2012,Gopalakrishnan2013,Li2013,Han2015,Li2017}. It was found that the miscibility-immiscibility phase transition can also be controlled by means of the SO coupling in the two-component BECs \cite{Lin2011}. Subsequently, the phase separation in the SO coupled BECs in quasi-one dimensional trap has been discussed \cite{Gautam2014}. However, so far it is not clear for the entire phase diagram of the existence of phase separation in SO coupled spin-1 BECs.

The rotation is an important experimental method of creating vortices and vortex lattices in the BECs \cite{Raman2001,Schweikhard2004,Williams2010,Martikainen2002,Mizushima2004}.
If the BECs in double-well potential are rotated, it was shown that the hidden vortices can occur distributed along
the central barrier \cite{Wen2010}. These hidden vortices have no visible cores in density
distributions but have phase singularities in phase distributions. In addition, they have an important characteristic
which can carry angular momentum like the usual vortices \cite{Wen2010}. Later, the pinning of hidden vortices in the BECs with a rotating double-well
potential and corotating optical lattice have been studied \cite{Mithun2014}. The vortices with hidden angular momentum have also been introduced \cite{Weiner2017}. Previous works on the hidden vortices have been limited to a single-component BECs. To the best of our
knowledge, it remains an open question that how to obtain the hidden vortices in spin-1 BECs so far, which is eager to be
explored.

In this paper, we address the question of the occurrence of the phase separation and hidden vortices in the rotating SO coupled spin-1 BECs.
We mainly concentrate on the phase separation in the SO coupled spin-1 BECs without the rotation. It is shown that the SO coupling is the key to
realize the phase separation in the present system. More remarkably, we determine the entire phase diagram of the existence of
phase separation in SO coupled spin-1 BECs, as compared with the previous work \cite{Gautam2014}. In the presence of the rotation, we
further demonstrate that the SO coupling can lead to the hidden vortices and hidden vortex-antivortex pairs, which provides a new method
of creating the hidden vortices in the BECs. Furthermore, we also determine the entire phase diagram of these two vortex states, depending on the interplay of
the SO coupling strength, the rotation frequency, and Rabi frequency, which shows the critical condition and phase boundary of
the occurrence of such two vortex states. The hidden vortices here are
proved to be long-lived in the time scale of experiment.

The paper is organized as follows. In Sec. II we formulate
the model Hamiltonian describing the rotating spin-1 BECs with laser-induced SO
coupling and Rabi coupling, and briefly introduce the numerical method. In Sec. III we mainly
discuss the phase separation in the SO coupled spin-1 BECs without the rotation. In Sec. IV we further
demonstrate that the SO coupling can lead to the hidden vortices and hidden vortex-antivortex pairs
in the presence of the rotation. The entire phase diagram of these two vortex states also be determined. Finally, we
conclude the main results of the work in Sec. V.

\section{MODEL HAMILTONIAN AND METHOD}
We consider the two-dimensional
ferromagnetic BECs with a laser-induced SO
coupling \cite{Lan2014,Gautam2015,Cheng2014,Salasnich2013,Kasamatsu2015,Zhang2013} in a frame rotating with frequency $\Omega$ \cite{Liu2012,Xu2011}.
In the mean-field approximation, the Hamiltonian is written
as \cite{Ji2008,Liu2012,Gautam2014,Martikainen2002,Mizushima2004}
%\begin{small}
\begin{eqnarray}
\begin{split}
H\!=\!&\int d\mathbf{r}\bigg\{\mathbf{\Psi}^\dagger\left(T\!+\!V_{H}(\mathbf{r})
\!-\!\Omega L_{z}
\!+\!\gamma P_{x}F_{z}\!+\!\Omega_{R}F_{x}\right)\mathbf{\Psi} \\
&\!+\!\left(\frac{c_{0}}{2}n^{2}\!+\!\frac{c_{2}}{2}[(n_{1}\!-\!n_{-1})^{2}
\!+\!2|\Psi^{\ast}_{1}\Psi_{0}\!+\!\Psi^{\ast}_{0}\Psi_{-1}|^{2}]\right)\bigg\},
\end{split}
\end{eqnarray}
%\end{small}
where $\mathbf{\Psi} =[\Psi_{1}(\mathbf{r}),\Psi_{0}(\mathbf{r}),\Psi_{-1}(\mathbf{r})]^{T}$
is the order parameter of the BECs with normalization
$\int d\mathbf{r}\mathbf{\Psi}^{\dagger}\mathbf{\Psi}=N$, and $N$ is the
total particle number. The total particle density is defined by $n=\sum_{\mathrm{m}}n_{\mathrm{m}}$,
wherein $n_{\mathrm{m}}=|\Psi_{\mathrm{m}}(\mathbf{r})|^{2}$ with
$\mathrm{m}=0, \pm1$. The kinetic-energy term $T=-\hbar^{2}\nabla^{2}/(2m)$,
where $m$ is the mass of a $^{87}$Rb atom and $\hbar$ is the Planck constant.
For simplicity, we assume that the harmonic trapping
frequencies satisfy $\omega_{z}\gg\omega_{\bot}$, where $\omega_{\bot}=2\pi\times40$ Hz and $\omega_z=2\pi\times800$ Hz are
the radial and axial trapping frequencies.
Then, the condensates are
pressed into a pancake so that the system
is effectively two dimensional. The two-dimensional confinement potential
$V_{H}(\mathbf{r})=m[\omega^{2}_{\bot}(x^{2}+y^{2})]/2$.
The projection of the angular momentum to the $z$ axis $L_{z}=-i\hbar(x\partial_{y}-y\partial_{x})$.
The fourth term of the Hamiltonian represents the one-dimensional SO coupling along the
$x$ direction, in which $p_{x}=-i\hbar\partial_{x}$ is the momentum operator along the $x$
direction, $\gamma$ is the SO coupling strength which can be controlled by the wavelength of the Raman
lasers and the angle between Raman beams in experiment \cite{Lin2011}. $\Omega_{R}$ is the Rabi frequency. $F_{z}$
and $F_{x}$ are the $3\times3$ Pauli matrices of the spin angular momentum operators in the $z$ and $x$
directions, which can be expressed as

\begin{equation}
F_z={
\left( \begin{array}{ccc}
1 & 0 & 0\\
0 & 0 & 0\\
0 & 0 & -1
\end{array}
\right )},
F_x=\frac{1}{\sqrt{2}}{
\left( \begin{array}{ccc}
0 & 1 & 0\\
1 & 0 & 1\\
0 & 1 & 0
\end{array}
\right ).}
\end{equation}
For the interaction terms, the coupling parameters
are given by $c_{0} = 4\pi\hbar^{2}(a_{0}+2a_{2})/3m$ and $c_{2} = 4\pi\hbar^{2}(a_{2}-a_{0})/3m$,
where $a_{0,2}$ are two-body s-wave
scattering lengths for total spin $0,2$. The wave functions of spin-1 BECs are formulated as the dimensionless
coupled Gross-Pitaevskii equations \cite{Ji2008,Liu2012,Gautam2014,Martikainen2002,Mizushima2004}
\begin{eqnarray}
i\frac{\partial\psi_{1}}{\partial t}&=&[-\frac{1}{2}\nabla^{2}+V+\lambda_{0}\rho+\lambda_{2}(\rho_{1}+\rho_{0}-\rho_{-1}) \nonumber\\
&&+i\tilde{\Omega}(x\partial_{y}-y\partial_{x})]\psi_{1}-i\kappa\frac{\partial\psi_{1}}{\partial x}+\frac{\tilde{\Omega}_{R}}{\sqrt{2}}\psi_{0} \nonumber\\
&&+\lambda_{2}\psi^{\ast}_{-1}\psi_{0}^{2},
\end{eqnarray}
\begin{eqnarray}
i\frac{\partial\psi_{0}}{\partial t}&=&[-\frac{1}{2}\nabla^{2}+V+\lambda_{0}\rho+\lambda_{2}(\rho_{1}+\rho_{-1}) \nonumber\\
&&+i\tilde{\Omega}(x\partial_{y}-y\partial_{x})]\psi_{0}+\frac{\tilde{\Omega}_{R}}{\sqrt{2}}(\psi_{1}+\psi_{-1}) \nonumber\\
&&+2\lambda_{2}\psi_{1}\psi_{-1}\psi_{0}^{\ast},
\end{eqnarray}
\begin{eqnarray}
i\frac{\partial\psi_{-1}}{\partial t}&=&[-\frac{1}{2}\nabla^{2}+V+\lambda_{0}\rho+\lambda_{2}(\rho_{0}+\rho_{-1}-\rho_{1}) \nonumber\\
&&+i\tilde{\Omega}(x\partial_{y}-y\partial_{x})]\psi_{-1}+i\kappa\frac{\partial\psi_{-1}}{\partial x}+\frac{\tilde{\Omega}_{R}}{\sqrt{2}}\psi_{0} \nonumber\\
&&+\lambda_{2}\psi_{1}^{\ast}\psi_{0}^{2},
\end{eqnarray}
where the dimensionless wave function $\psi_{j}=N^{-1/2}a_{h}\Psi_{j}$ and
the total condensate density $\rho=\rho_{1}+\rho_{0}+\rho_{-1}$ with $\rho_{j}=|\psi_{j}|^{2}$ $(j=1,0,-1)$.
The dimensionless optical trapping potential $V=(x^{2}+y^{2})/2$. The dimensionless interaction strengths
$\lambda_{0}=4\pi N(a_{0}+2a_{2})/3a_{h}$ and $\lambda_{2}=4\pi N(a_{2}-a_{0})/3a_{h}$.
We choose $a_{2}=(100.4\pm0.1)a_{B}$ for
total spin channel $F_{total}=2$ and $a_{0}=(101.8\pm0.2)a_{B}$ for total
spin channel $F_{total}=0$ \cite{Kempen2002,Stamper-Kurn2013}, where $a_{B}$ is
the Bohr radius. The characteristic length of the harmonic trap is
defined as $a_{h}=\sqrt{\hbar/m\omega_{\bot}}$. We can define the dimensionless rotation frequency
as $\tilde{\Omega}=\Omega/\omega_{\bot}$, the dimensionless strength of the SO coupling
as $\kappa=\gamma/\sqrt{\hbar\omega_{\bot}/m}$, and
the dimensionless Rabi frequency as $\tilde{\Omega}_{R}=\Omega_{R}/(\hbar\omega_{\bot})$. The time and the energy are
scaled in units of $\omega_{\bot}^{-1}$ and $\hbar\omega_{\bot}$.

\begin{figure}
\includegraphics[width= 0.48\textwidth]{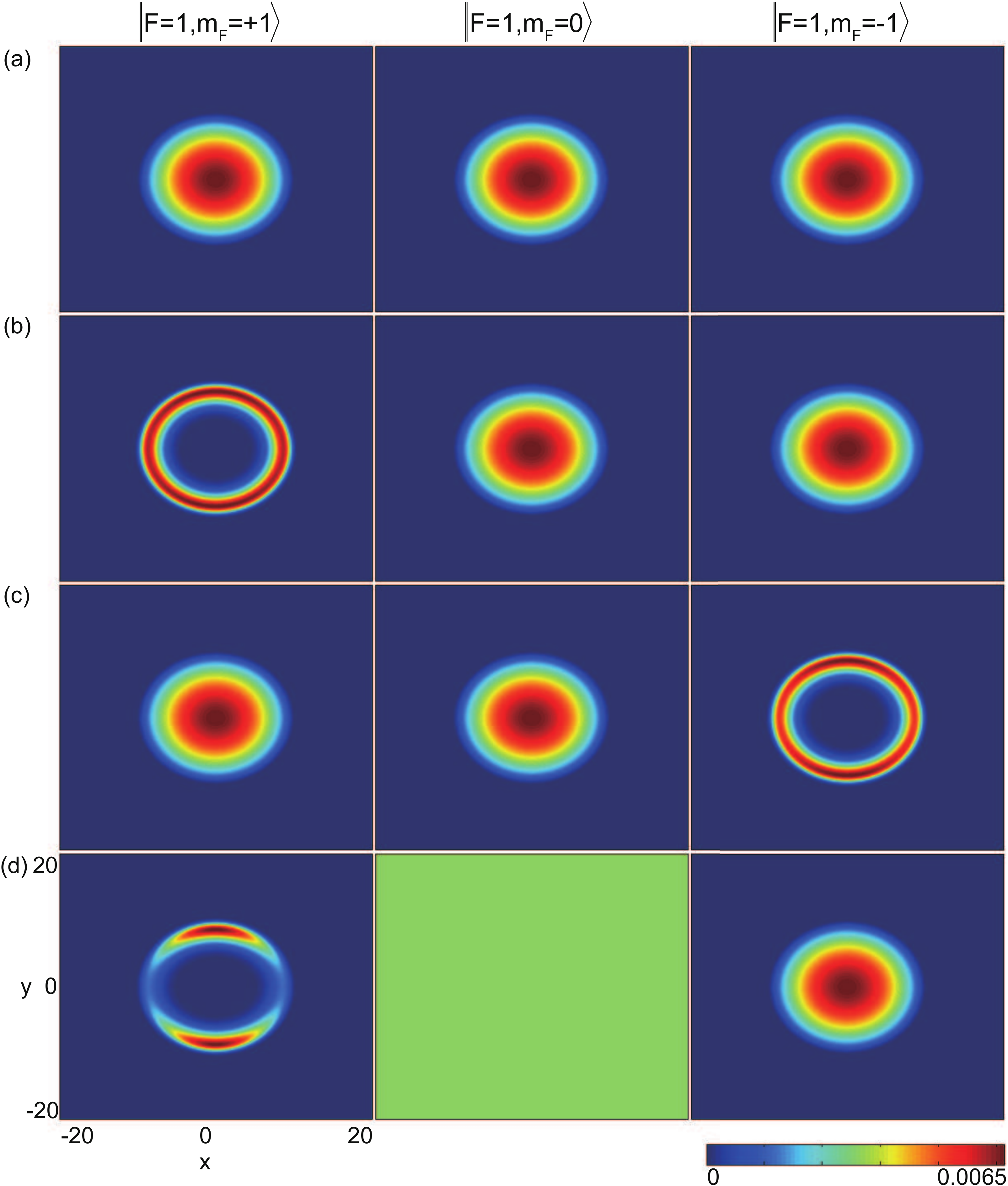}
\caption{(Color online). Phase separation of the ferromagnetic BECs induced by the SO coupling. (a)-(d) Particle density distributions of the ferromagnetic BECs for $\kappa=0$, 1.2, 2, and 6, respectively. The simulation is conducted by using the dimensionless spin-dependent interaction parameter $\lambda_{2}\!=\!-75$, spin-independent interaction parameter $\lambda_{0}\!=\!7500$, the rotation frequency $\widetilde{\Omega}=0$, the Rabi frequency $\tilde{\Omega}_{R}=0$, and the optical trap $\omega_{\bot}\!=\!2\pi\times40$Hz.} \label{Figure1}
\end{figure}

The stationary state of the system is obtained by using the standard
imaginary-time propagation combined with finite-difference methods \cite{Bao2004,Dalfovo1996}.
Equations (3)-(5) are solved by using the second-order
centered finite-difference for the spatial discretization and
the backward/forward Euler schemes of the linear/nonlinear terms for the time discretization.
In all numerical simulations, the size of the computational grid is 400$\times$400, corresponding to the field of view
being 40$\times$40 ($a_{h}^{2}$) or $68\times68$ ($\mu m^2$).
We start with a trial Gaussian
wave function for the three components and propagate the wave function in
imaginary time. A sufficiently large number of time steps is chosen, which guarantees that we have
reached a steady state. The final steady state is also been checked by the normalized random number as the initial state,
which suggests that our final stationary state is independent of the initial trial wave function. The dynamic evolution of the spin-1 BECs is
obtained by using a time-splitting spectral method with the
time stepping being $(8\times10^{-4})/\omega_{\bot}$ \cite{Wang2007,Wangtwo2007,Li2012}.

\section{Phase separation induced by SO coupling}
We first consider $\tilde{\Omega}=0$ case. It is found that the SO coupling can induce phase separation of the ferromagnetic BECs. In Fig. 1 we show the densities of the ground state of the BECs with different SO coupling strengths. Without the SO coupling, the ground state is miscible and no phase separation occurs, as shown in Fig. 1(a). As the SO coupling strength increases, the ground state is immiscible between the $m_{F}\!=\!+1$ and $-1$ components, proving that phase separation has taken place, as depicted in Figs. 1(b)-1(d). When the SO coupling is sufficiently strong, there is almost no particle in the $m_{F}\!=\!0$ component and a maximum of phase separation emerges between the $m_{F}\!=\!+1$ and $-1$ components, as shown in Fig. 1(d). The effect of the Rabi frequency $\tilde{\Omega}_{R}$ on the phase separation is also studied. We take the state shown in Fig. 1(d) as an example, which is used to investigate the effect of the Rabi frequency. If the Rabi coupling is considered such as $\tilde{\Omega}_{R}=0.1$, there remains the phase separation between the $m_{F}\!=\!+1$ and $-1$ components. However, the particles in the $m_{F}\!=\!0$ component increase as compared with the Fig. 1(d), as reflected in Fig. 2(a). When the Rabi frequency is stronger, such as $\tilde{\Omega}_{R}=4$, the system favors phase miscibility, as presented in Fig. 2(b). Therefore, when both $\kappa$ and $\tilde{\Omega}_{R}$ are considered, the occurrence of phase separation depends on the competition between $\kappa$ and $\tilde{\Omega}_{R}$. To illustrate this, in Fig. 3 we plot the
ground-state phase diagram by solving Equations (3)-(5) for a large number of $\kappa$ and $\tilde{\Omega}_{R}$ values.
When $\tilde{\Omega}_{R}$ is lager than a critical value $\tilde{\Omega}_{R(c)}=1.8$, the system manifests the phase miscibility for all $\kappa$ value in the present calculations. For $\tilde{\Omega}_{R}\leq1.8$, we observe the phase separation through tuning the strength of the SO coupling. We also give a critical value of producing phase separation with $\kappa_{c}=0.4$. Figure. 3 presents the critical condition of the occurrence of the phase separation in SO coupled spin-1 BECs, which offers a clear illustration and entire phase diagram on the phase separation, as compared with the previous work \cite{Gautam2014}.

\begin{figure}
\includegraphics[width= 0.48\textwidth]{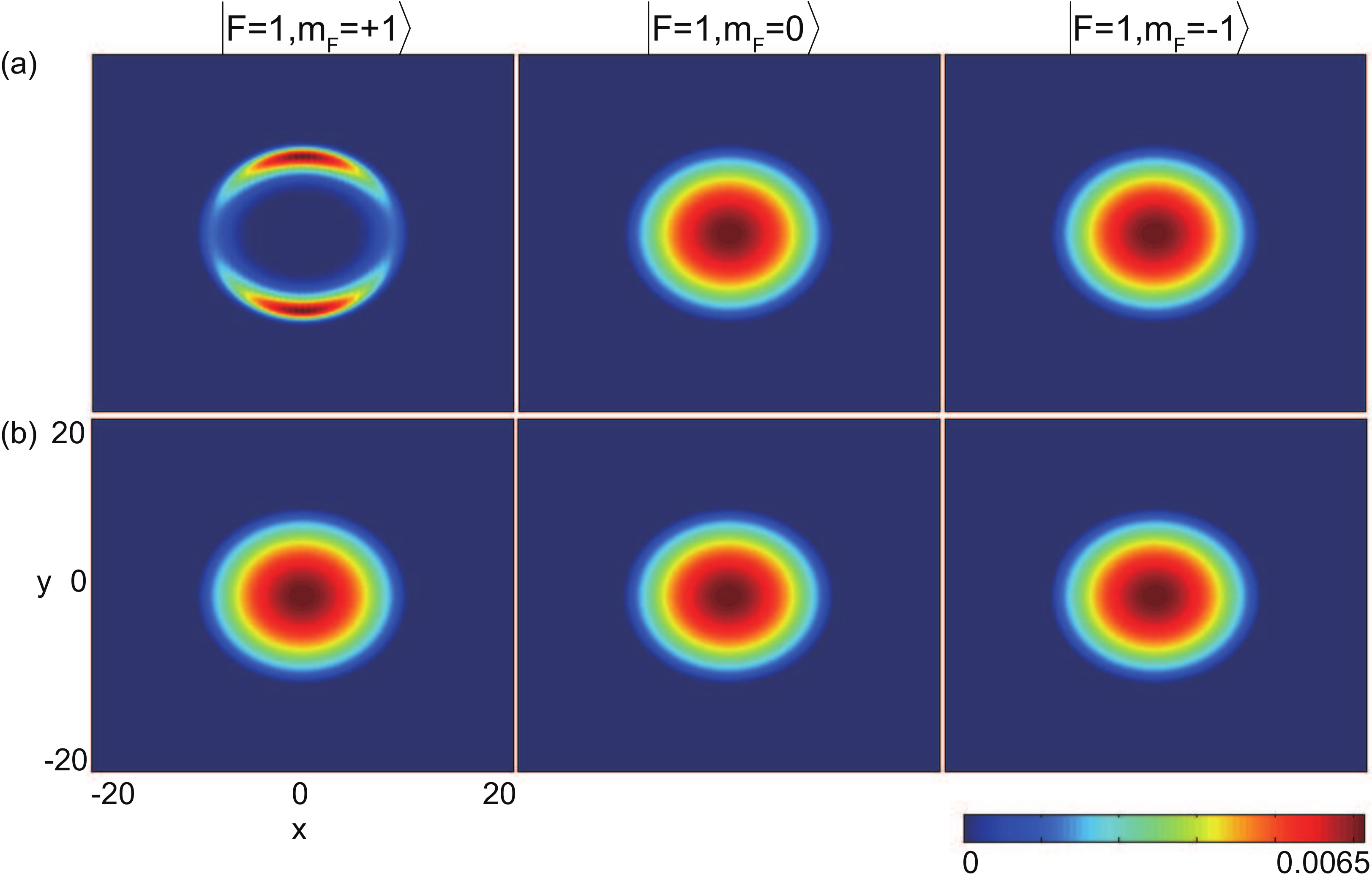}
\caption{(Color online). Influence of the Rabi frequency $\widetilde{\Omega}_{R}$ on the phase separation. (a) and (b) Show particle density distributions of the ferromagnetic BECs for $\widetilde{\Omega}_{R}=0.1$ and 4, respectively. The simulation uses $\kappa=6$ with the other parameters being the same as ones in Fig. 1.} \label{Figure2}
\end{figure}

\begin{figure}
\includegraphics[width= 0.46 \textwidth]{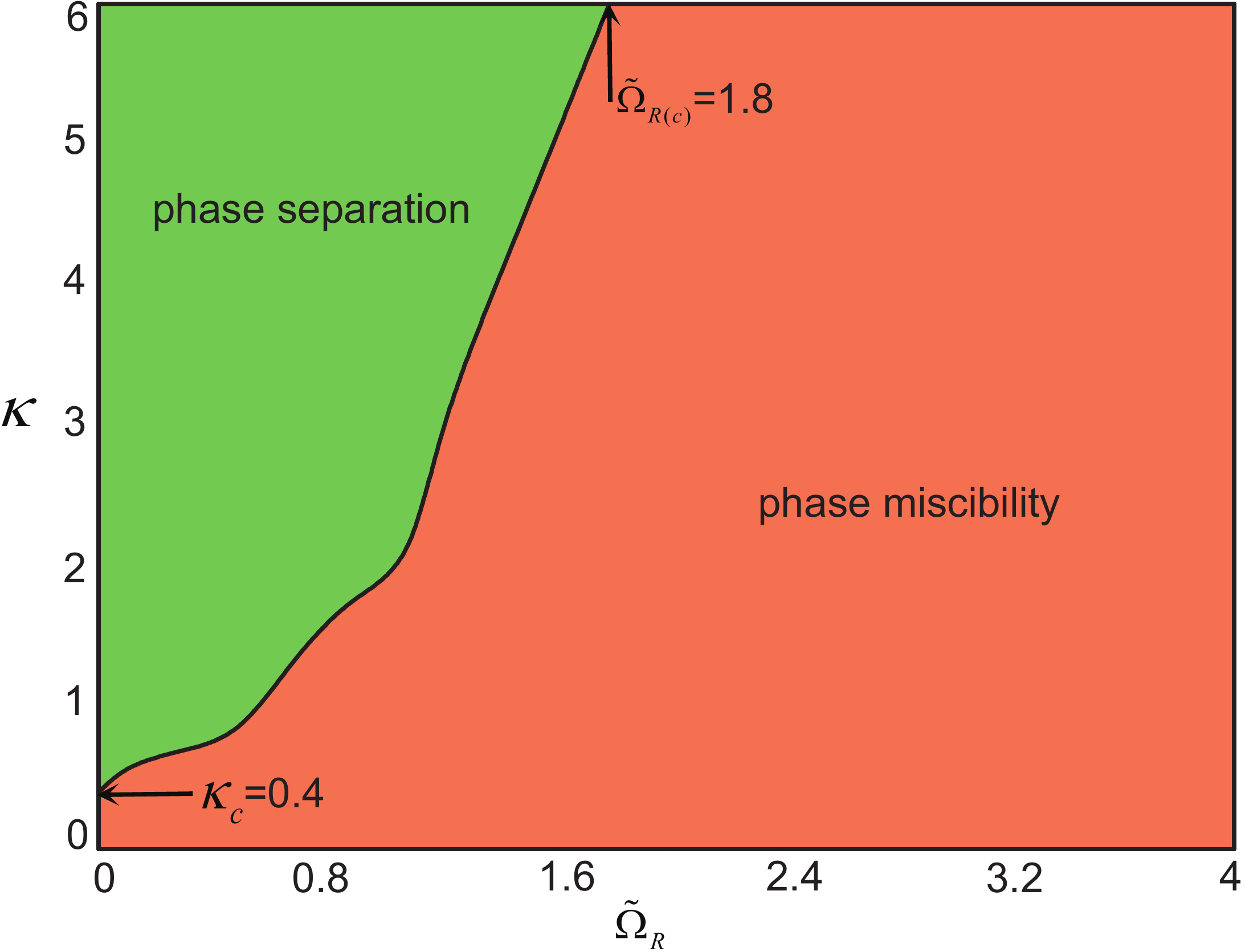}
\caption{(Color online). Phase diagram of the ferromagnetic BECs as functions of $\widetilde{\Omega}_{R}$ and $\kappa$ in the absence of the rotation.
The remaining parameters are the same as ones in Fig. 1.} \label{Figure3}
\end{figure}

\begin{figure}
\includegraphics[width= 0.48\textwidth]{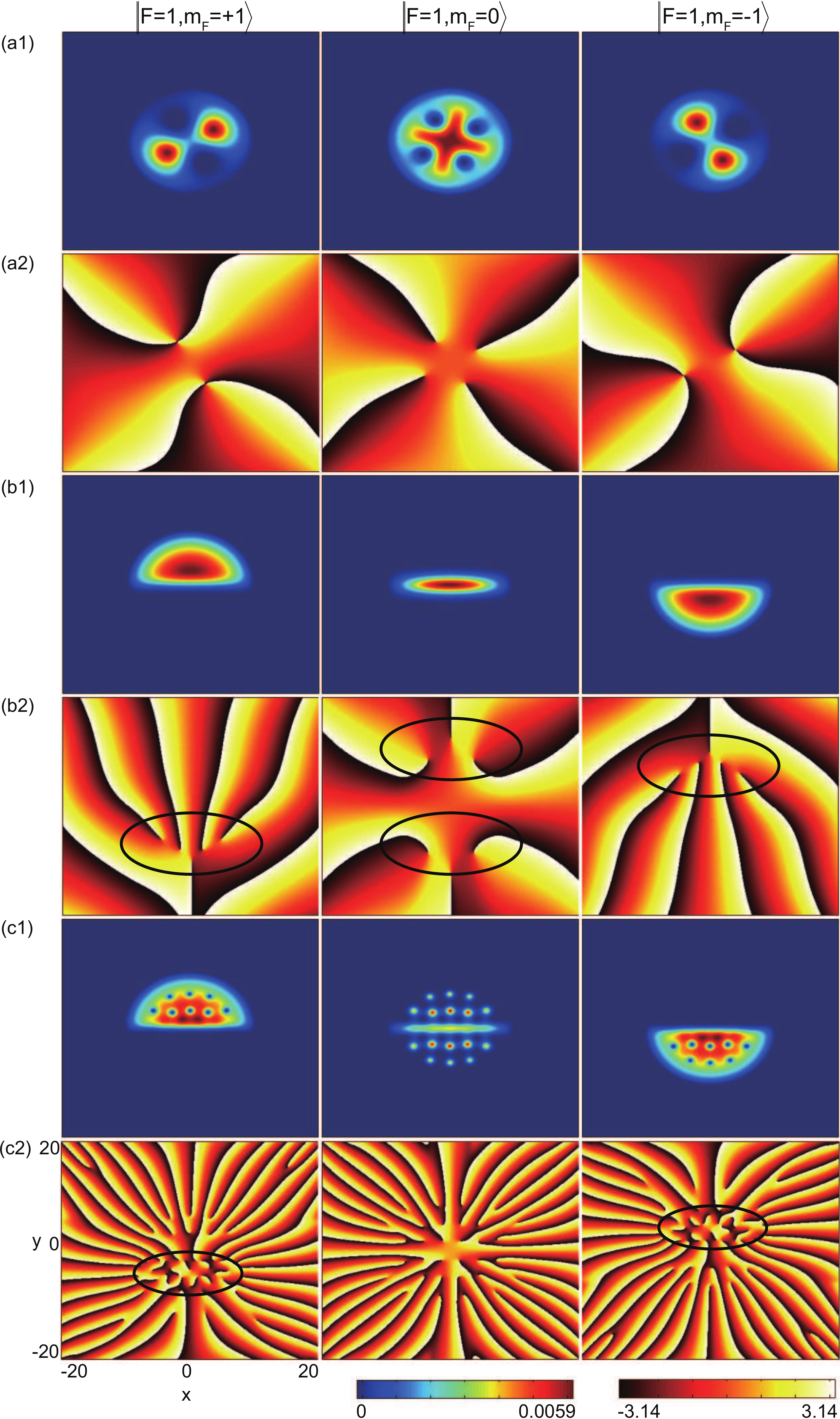}
\caption{(Color online). Hidden vortices and hidden vortex-antivortex pairs induced by the SO coupling. (a1) and (a2) Particle densities and phase distributions of the  visible vortices with $\kappa=0$, $\tilde{\Omega}_{R}=0$ and $\tilde{\Omega}=0.1$. (b1) and (b2) Particle densities and phase distributions of the hidden vortices with $\kappa=0.5$, $\tilde{\Omega}_{R}=0$ and $\tilde{\Omega}=0.1$. The position of the hidden vortex is highlighted by the black ellipse. (c1) and (c2) Particle densities and phase distributions of the hidden vortex-antivortex pairs with $\kappa=0.5$, $\tilde{\Omega}_{R}=0$ and $\tilde{\Omega}=0.4$. The position of vortex-antivortex pair is highlighted by the black ellipse. The remaining parameters are the same as ones
in Fig.1.} \label{Figure4}
\end{figure}

\begin{figure}
\includegraphics[width= 0.49\textwidth]{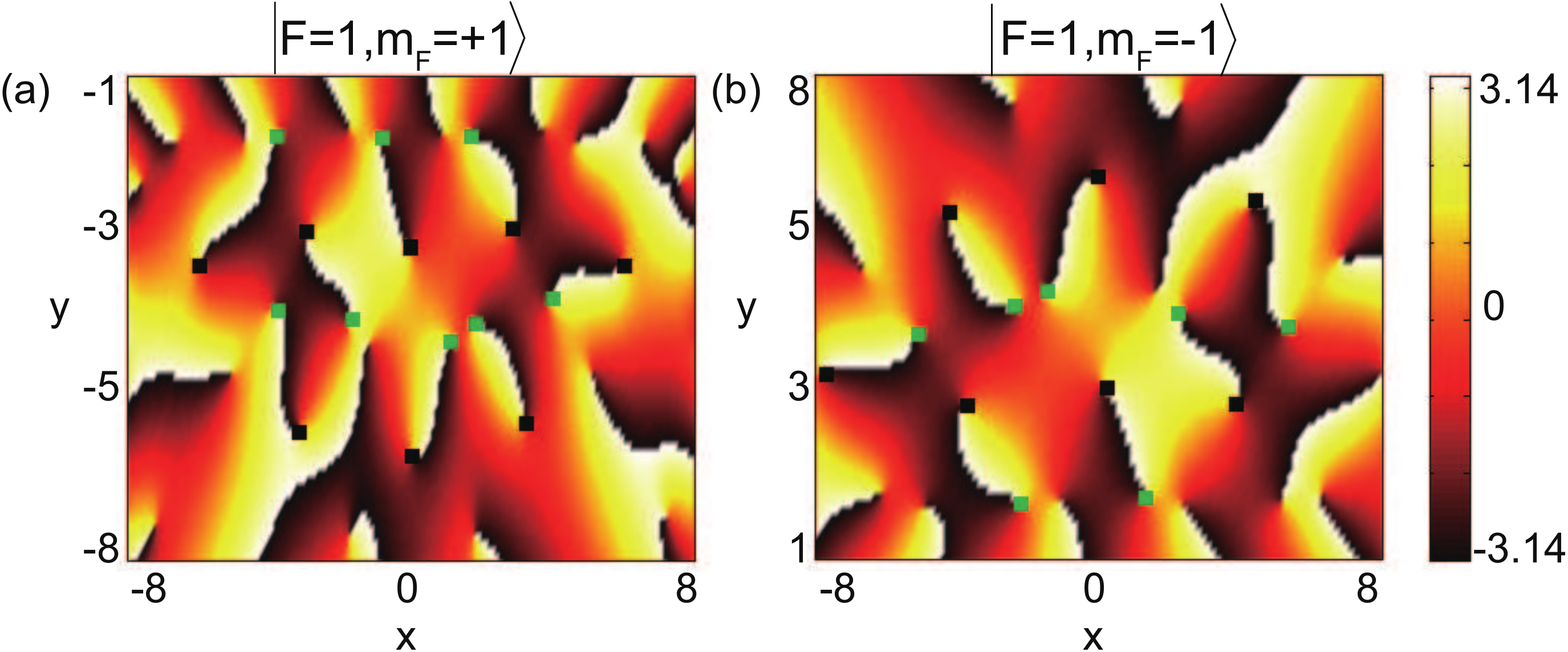}
\caption{(Color online). (a) and (b) Local enlargement of the vortex-antivortex
pair in Fig. 4(c2). The black square and the green square mark the position of
the vortex-antivortex
pair.} \label{Figure5}
\end{figure}

\begin{figure}
\includegraphics[width= 0.49\textwidth]{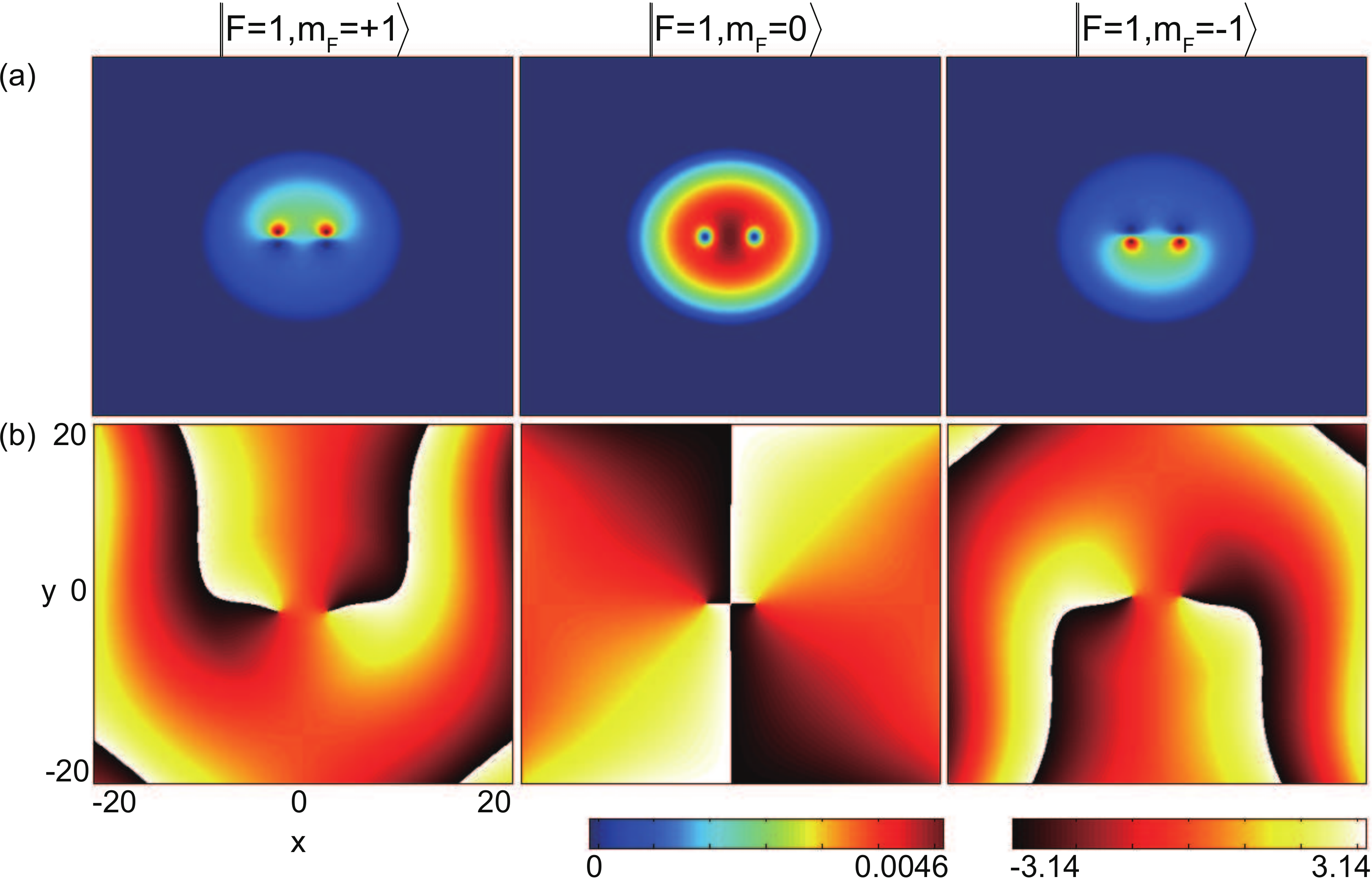}
\caption{(Color online). Influence of the Rabi frequency $\widetilde{\Omega}_{R}$ on the hidden vortices corresponding to Figs. 4(b1) and 4(b2). (a) and (b) Particle density and phase distributions of the ferromagnetic BECs in the presence of $\widetilde{\Omega}_{R}$. The simulation uses $\widetilde{\Omega}_{R}=1$ with the other parameters being the same as ones in Figs. 4(b1) and 4(b2).} \label{Figure6}
\end{figure}

\begin{figure}
\includegraphics[width= 0.46\textwidth]{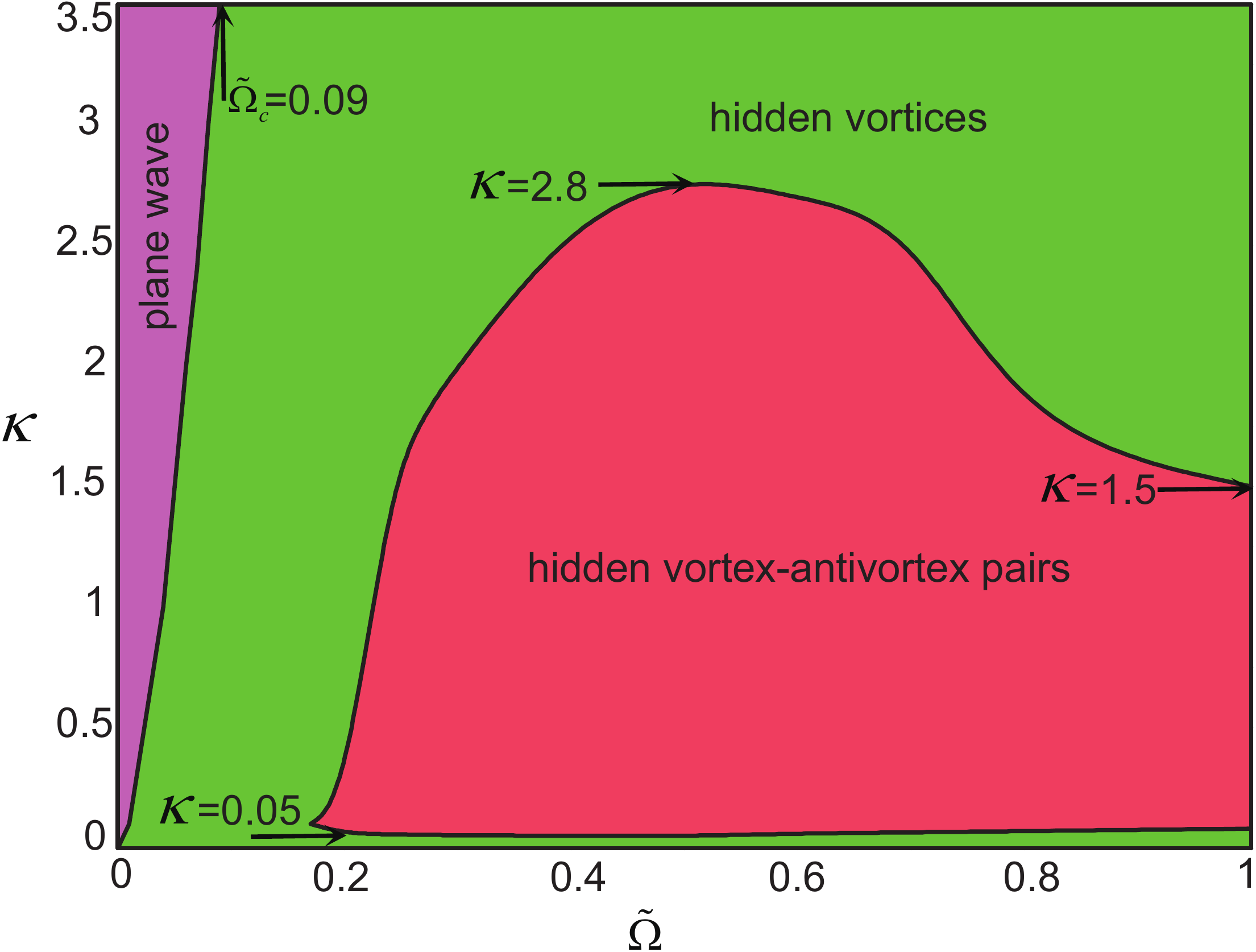}
\caption{(Color online). Phase diagram of the ferromagnetic BECs as functions of $\kappa$ and $\tilde{\Omega}$ without the Rabi coupling. The other parameters are the same as ones in Fig. 1.} \label{Figure7}
\end{figure}

\begin{figure}
\includegraphics[width= 0.46\textwidth]{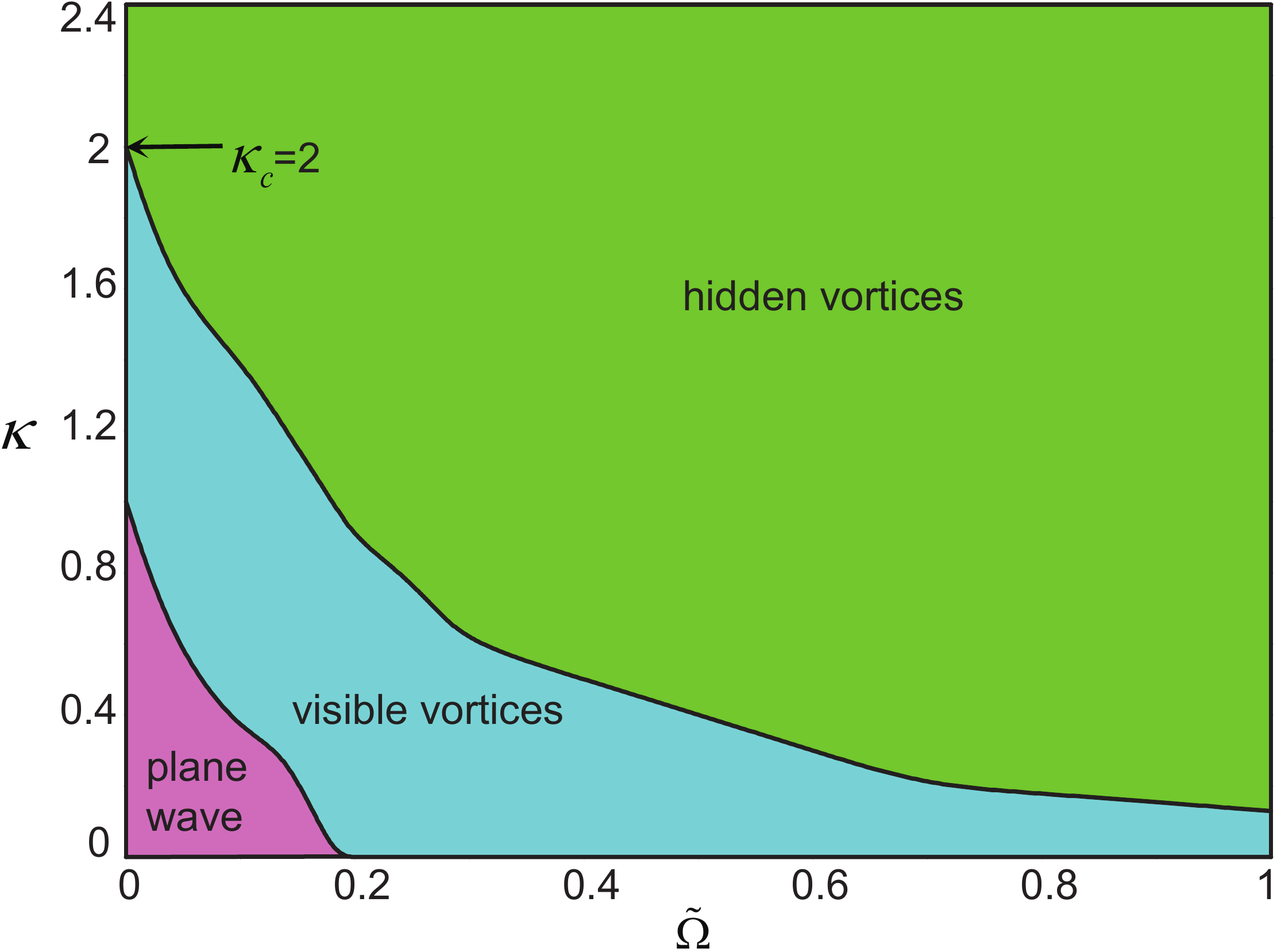}
\caption{(Color online). Phase diagram of the ferromagnetic BECs as functions of $\kappa$ and $\tilde{\Omega}$ with a fixed Rabi frequency $\widetilde{\Omega}_{R}=1$. The other parameters are the same as ones in Fig. 1.} \label{Figure8}
\end{figure}

\section{Hidden vortices and hidden vortex-antivortex pairs induced by SO coupling}
Next we investigate $\tilde{\Omega}\neq0$ case. If the system is only subject to the rotation, the ground state of the BECs is similar to a coreless vortex \cite{Martikainen2002}. The corresponding density and phase distributions are shown in Figs. 4(a1) and 4(a2). In this case, we can see clearly vortex cores from the density distributions. The corresponding phase singularities can also be observed from the phase distributions, and such these vortices are known as visible vortices. Subsequently, in order to study the effect of the SO coupling, $\tilde{\Omega}_{R}$ is still taken as zero. When $\kappa=0.5$ and $\tilde{\Omega}=0.1$, the hidden vortices are found in the present system in addition to the phase separation, as shown in Figs. 4(b1) and 4(b2). Form the density distributions we can see that the $m_{F}\!=\!+1$ and $-1$ components becomes spatially separated along $y$ direction. In addition, there are no vortex cores in density profile. However, we can observe the phase singularities of three spin components from the phase profile, as highlighted by the black ellipses in Fig. 4(b2). Such the ground-state structures can be called as the hidden vortices. Since the hidden vortices have an important feature which can carry angular momentum like the visible vortices \cite{Wen2010,Mithun2014}, we further confirm that the hidden vortices here carry angular momentum by calculating the mean angular momentum per atom. Figs. 4(b1) and 4(b2) confirm that the SO coupling can result in the hidden vortices in the presence of slow rotation. Based on Figs. 4(b1) and 4(b2), if the rotation frequency is further enhanced, it is found that the SO coupling can induce the emergence of hidden vortex-antivortex pairs. The corresponding result are presented in Figs. 4(c1) and 4(c2), at which these hidden vortex-antivortex pairs are highlighted by the black ellipses in phase profile. Meanwhile, due to the increasing $\tilde{\Omega}$, there are also some visible vortices in density profile, as shown in Fig. 4(c1). Local enlargements of these hidden vortex-antivortex pairs are shown in Fig. 5. We next investigate the effect of the Rabi frequency on the hidden vortices. In Fig. 6, we consider the Rabi frequency, for example $\tilde{\Omega}_{R}=1$, as compared with the case in Figs. 4(b1) and 4(b2). The result shows that the visible vortices appear in the system. We can see the vortex cores from density profile, as shown in Fig. 6(a). At the same time, the corresponding phase singularities can also be observed in the phase profile, as shown in Fig. 6(b). Figure 6 proves that the increase of the Rabi frequency can cause the transformation of a hidden vortex state to a visible vortex state in the rotating ferromagnetic BECs.

From Fig. 4 we can suppose that the occurrence of the hidden vortices and hidden vortex-antivortex pairs depends on the interplay of the SO coupling strength and the rotation frequency when the Rabi frequency is zero. To further confirm this, we plot the phase diagram as a function of $\kappa$ and $\tilde{\Omega}$ in Fig. 7. When
$\tilde{\Omega}$ is less than a critical value $\tilde{\Omega}_{c}=0.09$, the system is found to translate from the plane wave phase to the hidden vortices as $\tilde{\Omega}$ increases. When $\tilde{\Omega}\geq\tilde{\Omega}_{c}$, the system only support the hidden vortices for the very weak or strong SO coupling strength, such as $\kappa<0.05$ or $2.8<\kappa\leq3.5$. For the weak SO coupling strength like $0.05\leq\kappa<1.5$, increasing $\tilde{\Omega}$ can induce the phase transition from the hidden vortices to the hidden vortex-antivortex pairs. It should be noted that, for the moderate strength of the SO coupling such as $1.5\leq\kappa\leq2.8$, as $\tilde{\Omega}$ is increased, the ground state structures undergo phase transitions from the hidden vortices to the hidden vortex-antivortex pairs, and eventually develop the hidden vortices. Figure. 7 presents the critical condition of creating the hidden vortices and vortex-antivortex pairs in the SO coupled BECs in the absence of the Rabi frequency. In Fig. 8 we further investigate the phase diagram as a function of $\kappa$ and $\tilde{\Omega}$ when the Rabi frequency is nonzero. As the SO coupling strength and the rotation frequency are increased, the system undergoes phase transitions from the plane wave to the visible vortices, and eventually develop the hidden vortices. Especially, when $\kappa$ is lager than a critical value $\kappa_{c}=2$, the ground state is the hidden vortices irrespective of the rotation frequency. Figure. 8 confirms that, for a given Rabi frequency, the hidden vortices are easier to occur when both the SO coupling strength and the rotation frequency are strong.

So far we have clarified the phase diagram at the fixed Rabi frequency. In Fig. 9 we further give the phase diagram as a function of $\tilde{\Omega}_{R}$ and $\tilde{\Omega}$ for the given SO coupling like $\kappa=0.5$. When $\tilde{\Omega}$ is less than a critical value $\tilde{\Omega}_{c}=0.2$, with the increasing Rabi frequency, the system undergoes phase transitions from the hidden vortices to the visible vortices, and eventually develop the plane wave. It should be noticed that the hidden vortex-antivortex pairs can occur in a very narrow region of parameters. In the rest region of parameters, increasing the Rabi frequency can transform the hidden vortices to the visible vortices for the fixed rotation frequency. We also emphasize that the system only exhibit the plane wave for any Rabi frequency when $\tilde{\Omega}<0.01$. Figure. 9 shows that, for the given SO coupling strength, the hidden vortices are easier to occur when the rotation frequency are strong or the Rabi frequency is weak.

Finally, we simulate the dynamic evolution of the hidden vortices. We use the hidden vortex state shown in Figs. 4(b1) and 4(b2) as initial state of the dynamic evolution. Figure. 10 presents the evolution of the hidden vortices versus time. In Figs. 10(a1) and 10(a2) we show the morphology of the hidden vortices when $t=16$ ms. It is shown that there is only the deflection of the density profile as compared with the Fig. 4(a1). From the phase profile we can see the phase singularities corresponding to the hidden vortices [see Fig. 10(a2)]. As the time evolution, when $t=32$ ms, the density profile is further deflected. It can be seen that the hidden vortices are well kept in the system [see Fig. 10(b2)]. The hidden vortices remain existing for a longer time like at $t=64$ ms in addition to the deflection of the density profile [see Fig. 10(c2)]. Figure 10 indicates that the hidden vortices have long lifetimes that are even beyond the time window of our simulations. We can expect that these hidden vortices are able to exist in an atomic gas as a long-lived configuration.

\begin{figure}
\includegraphics[width= 0.46\textwidth]{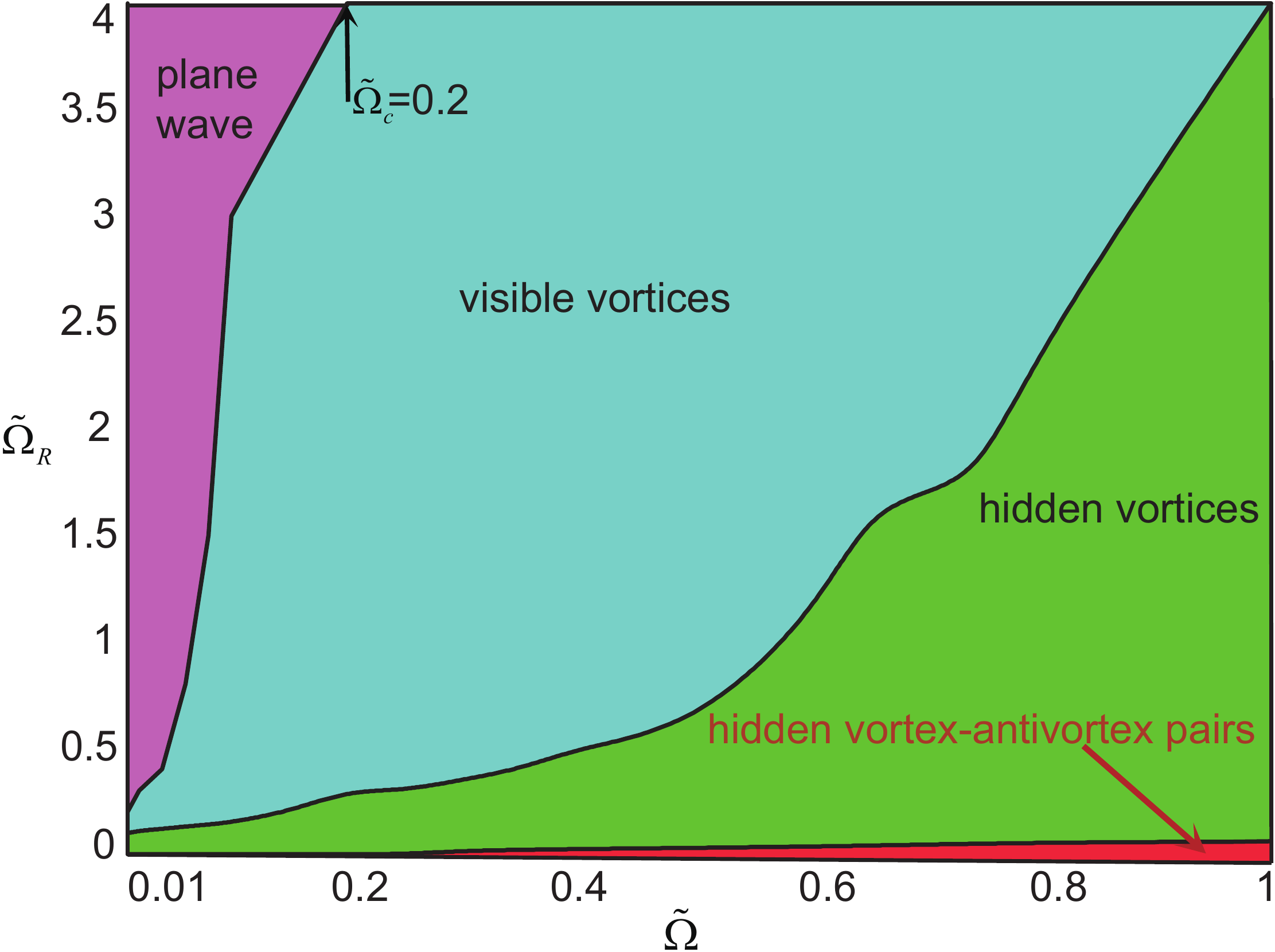}
\caption{(Color online). Phase diagram of the ferromagnetic BECs as functions of $\widetilde{\Omega}_{R}$ and $\tilde{\Omega}$ with a given SO coupling strength $\kappa=0.5$. The other parameters are the same as ones in Fig. 1.} \label{Figure9}
\end{figure}

\begin{figure}
\includegraphics[width= 0.48\textwidth]{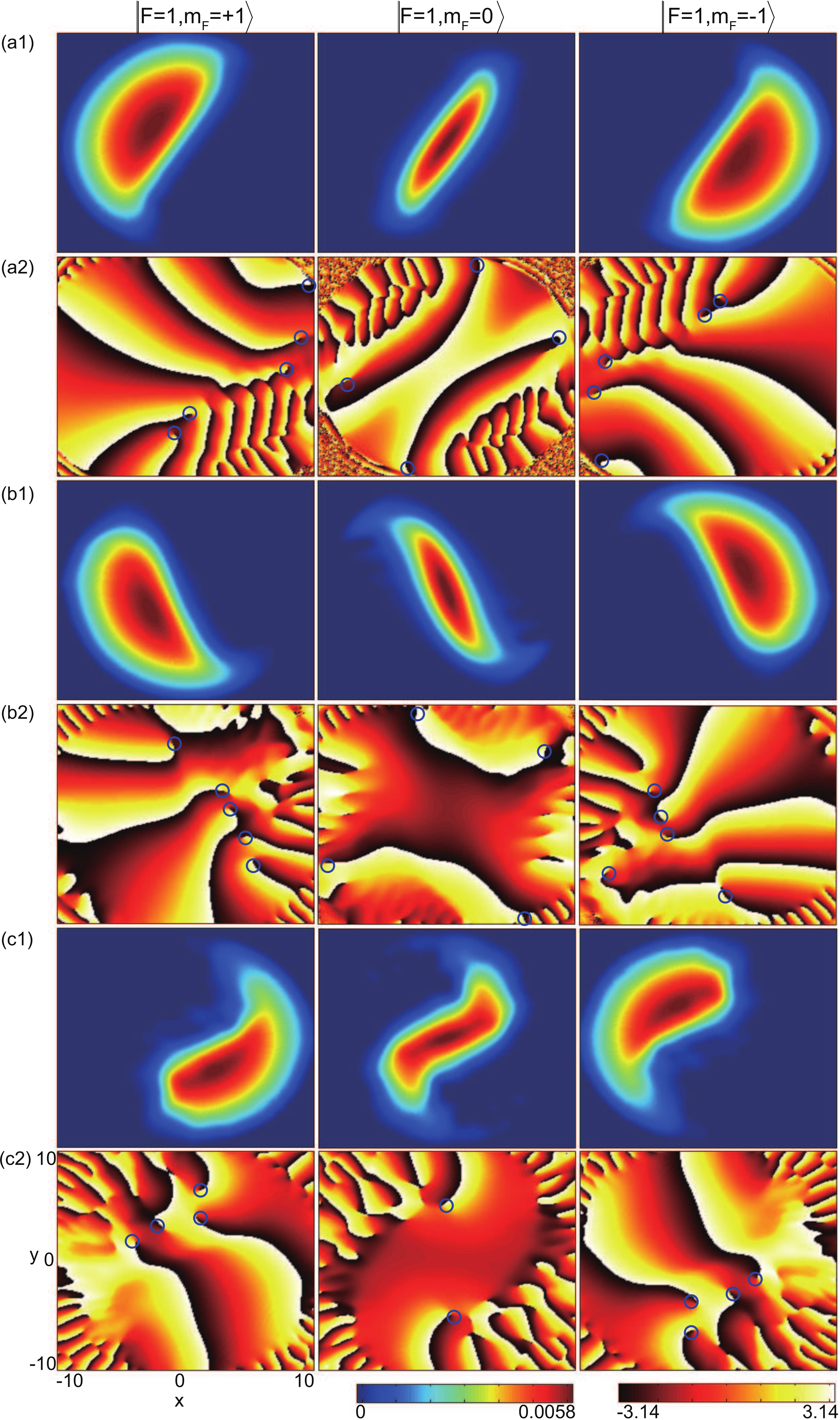}
\caption{(Color online). Real-time evolutions of the hidden vortices corresponding to the Figs. 4(b1) and 4(b2). (a1) and (a2) Particle density and phase distributions of the  ferromagnetic BECs when $t=16$ ms. (b1) and (b2) Particle density and phase distributions of the  ferromagnetic BECs when $t=32$ ms. (c1) and (c2) Particle density and phase distributions of the ferromagnetic BECs when $t=64$ ms. The position of the hidden vortex is highlighted by the blue circular ring. The simulation parameters are the same as the ones in Figs. 4(b1)-4(b2).} \label{Figure10}
\end{figure}

\section{CONCLUSION}
We have studied phase separation and hidden vortices in the laser-induced SO coupled ferromagnetic BECs with the rotation and Rabi coupling.
We have demonstrated that the SO coupling play a crucial role for the occurrence of
the phase separation in the present work. We have determined the corresponding entire phase diagram that indicates the critical
condition of obtaining the phase separation in SO coupled ferromagnetic BECs. We have found that the SO coupling can induce the hidden vortices and hidden vortex-antivortex pairs in the presence of the rotation.
We have predicted the rich phase diagrams of such two vortex states by changing the SO coupling strength, the rotation frequency, and Rabi frequency.
The dynamical stability of the hidden vortex have also been investigated.
Our study provides a method of creating the hidden vortices in spinor BECs, which paves the way for
future explorations of the hidden vortices in high-spin BECs, superfluid, and superconduction.
We expect that the present work will be useful for better
understanding of phase separation and hidden vortices in quantum systems.

\section*{ACKNOWLEDGMENT}
This work was supported by the NKRDP under grants Nos. 2016YFA0301500;
NSFC under grants No. 11434015, No. 61227902, No. 11611530676, and No. KZ201610005011;
SKLQOQOD under grants No. KF201403;
SPRPCAS under grants No. XDB01020300 and No. XDB21030300.


\begin{thebibliography}{99}
\bibitem{Stamper-Kurn1998}
D. M. Stamper-Kurn, M. R. Andrews, A. P. Chikkatur, S. Inouye, H.-J. Miesner, J. Stenger, and W. Ketterle,
Phys. Rev. Lett. \textbf{80}, 2027-2030 (1998).

\bibitem{Barrett2001}
M. D. Barrett, J. A. Sauer, and M. S. Chapman,
Phys. Rev. Lett. \textbf{87}, 010404 (2001).

\bibitem{Sadler2006}
L. E. Sadler, J. M. Higbie, S. R. Leslie, M. Vengalattore, and D. M. Stamper-Kurn,
Nature \textbf{443}, 312-315 (2006).

\bibitem{Ji2008}
A. C. Ji, W. M. Liu, J. L. Song, and F. Zhou,
Phys. Rev. Lett. \textbf{101}, 010402 (2008).

\bibitem{Khawaja2001}
U. A. Khawaja and H. Stoof,
Nature \textbf{411}, 918-920 (2001).

\bibitem{Choi2012}
J. Y. Choi, W. J. Kwon, and Y. I. Shin
Phys. Rev. Lett. \textbf{108}, 035301 (2012).

\bibitem{Kawaguchi2008}
Y. Kawaguchi, M. Nitta, and M. Ueda,
Phys. Rev. Lett. \textbf{100}, 180403 (2008).

\bibitem{Hall2016}
D. S. Hall, M. W. Ray, K. Tiurev, E. Ruokokoski, A. H. Gheorghe, and M. M\"{o}tt\"{o}nen,
Nat. Phys. \textbf{12}, 478 (2016).

\bibitem{Ray2014}
M. W. Ray, E. Ruokokoski, S. Kandel, M. M\"{o}tt\"{o}nen, and D. S. Hall,
Nature \textbf{505}, 657-660 (2014).


\bibitem{Ho1996}
T. L. Ho, and V. B. Shenoy,
Phys. Rev. Lett. \textbf{77}, 3276-3279 (1996).


\bibitem{Pu1998}
H. Pu and N. P. Bigelow,
Phys. Rev. Lett. \textbf{80}, 1130-1133 (1998).

\bibitem{Sabbatini2011}
J. Sabbatini, W. H. Zurek, and M. J. Davis,
Phys. Rev. Lett. \textbf{107}, 230402 (2011).

\bibitem{Zhou2008}
L. Zhou, J. Qian, H. Pu, W. P. Zhang, and H. Y. Ling,
Phys. Rev. A \textbf{78}, 053612 (2008).

\bibitem{Xi2011}
K. T. Xi, J. B. Li, and D. N. Shi,
Phys. Rev. A \textbf{84}, 013619 (2011).

\bibitem{Bandyopadhyay2017}
S. Bandyopadhyay, A. Roy, and D. Angom,
Phys. Rev. A \textbf{96}, 043603 (2017).

\bibitem{Myatt1997}
C. J. Myatt, E. A. Burt, R. W. Ghrist, E. A. Cornell, and C. E. Wieman,
Phys. Rev. Lett. \textbf{78}, 586-589 (1997).

\bibitem{Matuszewski2009}
M. Matuszewski, T. J. Alexander, and Y. S. Kivshar,
Phys. Rev. A \textbf{80}, 023602 (2009).

\bibitem{Matuszewski2010}
M. Matuszewski,
Phys. Rev. A \textbf{82}, 053630 (2010).

\bibitem{ocki2012}
T. \'{S}wis{\l}ocki and M. Matuszewski,
Phys. Rev. A \textbf{85}, 023601 (2012).

\bibitem{Lin2011}
Y. J. Lin, K. Jim\'{e}nez-Garc\'{l}a, and I. B. Spielman,
Nature \textbf{471}, 83-86 (2011).

\bibitem{Wu2016}
Z. Wu, L. Zhang, W. Sun, X. T. Xu, B. Z. Wang, S. C. Ji, Y. J. Deng, S. Chen, X. J. Liu, and J. W. Pan,
Science \textbf{354}, 83-88 (2016).

\bibitem{Huang2016}
L. H. Huang, Z. M. Meng, P. J. Wang, P. Peng, S. L. Zhang, L. C. Chen, D. H. Li, Q. Zhou, and J. Zhang,
Nat. Phys. \textbf{12}, 540 (2016).

\bibitem{Wang2010}
C. J. Wang, C. Gao, C. M. Jian, and H. Zhai,
Phys. Rev. Lett. \textbf{105}, 160403 (2010).

\bibitem{Su2012}
S.-W. Su, I.-K. Liu, Y.-C. Tsai, W. M. Liu, and S.-C. Gou,
Phys. Rev. A \textbf{86}, 023601 (2012).

\bibitem{Liu2012}
C. F. Liu and W. M. Liu,
Phys. Rev. A \textbf{86}, 033602 (2012).

\bibitem{Xu2011}
X. Q. Xu and J. H. Han,
Phys. Rev. Lett. \textbf{107}, 200401 (2011).

\bibitem{Sinha2011}
S. Sinha, R. Nath, and L. Santos,
Phys. Rev. Lett. \textbf{107}, 270401 (2011).

\bibitem{Hu2012}
H. Hu, B. Ramachandhran, H. Pu, and  X. J. Liu,
Phys. Rev. Lett. \textbf{108}, 010402 (2012).

\bibitem{Gopalakrishnan2013}
S. Gopalakrishnan, I. Martin, and E. A. Demler,
Phys. Rev. Lett. \textbf{111}, 185304 (2013).

\bibitem{Li2013}
Y. Li, G. I. Martone,  L. P. Pitaevskii, and S. Stringari,
Phys. Rev. Lett. \textbf{110}, 235302 (2013).

\bibitem{Han2015}
W. Han, G. Juzeli\={u}nas, W. Zhang, and W. M. Liu,
Phys. Rev. A \textbf{91}, 013607 (2015).

\bibitem{Li2017}
J. Li, Y. M. Yu, L. Zhuang, and W. M. Liu,
Phys. Rev. A \textbf{95}, 043633 (2017).

\bibitem{Gautam2014}
S. Gautam and S. K. Adhikari,
Phys. Rev. A \textbf{90}, 043619 (2014).

\bibitem{Raman2001}
C. Raman, J. R. Abo-Shaeer, J. M. Vogels, K. Xu, and W. Ketterle,
Phys. Rev. Lett. \textbf{87}, 210402 (2001).

\bibitem{Schweikhard2004}
V. Schweikhard, I. Coddington, P. Engels, S. Tung, and E. A. Cornell,
Phys. Rev. Lett. \textbf{93}, 210403 (2004).

\bibitem{Williams2010}
R. A. Williams, S. Al-Assam, and C. J. Foot,
Phys. Rev. Lett. \textbf{104}, 050404 (2010).

\bibitem{Martikainen2002}
J.-P. Martikainen, A. Collin, and K.-A. Suominen,
Phys. Rev. A \textbf{66}, 053604 (2002).

\bibitem{Mizushima2004}
T. Mizushima, N. Kobayashi, and K. Machida,
Phys. Rev. A \textbf{70}, 043613 (2004).

\bibitem{Wen2010}
L. H. Wen,  H. W. Xiong, and B. Wu,
Phys. Rev. A \textbf{82}, 053627 (2010).

\bibitem{Mithun2014}
T. Mithun, K. Porsezian, and B. Dey,
Phys. Rev. A \textbf{89}, 053625 (2014).

\bibitem{Weiner2017}
S. E. Weiner, M. C. Tsatsos, L. S. Cederbaum, and A. U. J. Lode,
Sci. Rep. \textbf{7}, 40122 (2017).

\bibitem{Lan2014}
Z. H. Lan and P. \"{O}hberg,
Phys. Rev. A \textbf{89}, 023630 (2014).

\bibitem{Gautam2015}
S. Gautam and S. K. Adhikari,
Phys. Rev. A \textbf{91}, 013624 (2015).

\bibitem{Cheng2014}
Y. S. Cheng, G. H. Tang, and S. K. Adhikari,
Phys. Rev. A \textbf{89}, 063602 (2014).

\bibitem{Salasnich2013}
L. Salasnich and B. A. Malomed,
Phys. Rev. A \textbf{87}, 063625 (2013).

\bibitem{Kasamatsu2015}
K. Kasamatsu,
Phys. Rev. A \textbf{92}, 063608 (2015).

\bibitem{Zhang2013}
Y. P. Zhang and C. W. Zhang,
Phys. Rev. A \textbf{87}, 023611 (2013).

\bibitem{Kempen2002}
E. G. M. van Kempen, S. J. J. M. F. Kokkelmans, D. J. Heinzen, and B. J. Verhaar,
Phys. Rev. Lett. \textbf{88}, 093201 (2002).

\bibitem{Stamper-Kurn2013}
D. M. Stamper-Kurn, M. Ueda,
Rev. Mod. Phys. \textbf{85}, 1191-1244 (2013).

\bibitem{Bao2004}
W. Z. Bao, Q. Du,
SIAM J. Sci. Comput. \textbf{25}, 1674-1697 (2004).

\bibitem{Dalfovo1996}
F. Dalfovo, S. Stringari,
Phys. Rev. A \textbf{53}, 2477-2485 (1996).

\bibitem{Wang2007}
H. Q. Wang,
International Journal of Computer Mathematics \textbf{84}, 925-944 (2007).

\bibitem{Wangtwo2007}
H. Q. Wang,
Journal of Computational and Applied Mathematics \textbf{205}, 88-104 (2007).

\bibitem{Li2012}
J. Li, D. S. Wang, Z. Y. Wu, Y. M. Yu, and W. M. Liu,
Phys. Rev. A \textbf{86}, 023628 (2012).


\end{thebibliography}
\end{document}